\documentclass[english,aps,pra,reprint,noshowpacs,superscriptaddress]{revtex4-1}   
\usepackage[T1]{fontenc}	% should generally be included for better accented-word behavior
\usepackage[latin9]{inputenc}	% should generally be included for better accent behavior
\usepackage{geometry}		% for controlling page margins
\usepackage{graphicx}
\usepackage[above,below]{placeins}	% allows use of \FloatBar­rier command to force section barriers
\usepackage{times}
\usepackage{hyperref}  
\hypersetup{colorlinks=true,urlcolor=blue,citecolor=blue,linkcolor=blue}   
\begin{document}

\title{Performance differences for in-class and online administration of low-stakes research-based assessments}

%\pacs{01.40.Fk, 01.40.G--, 01.40.gb I. }
\keywords{Concept Inventory, Computer Based Test, LASSO, Learning Assistant}

\author{Jayson M. Nissen}\affiliation{Department of Science Education, California State University Chico, Chico, CA, 95929, USA} 
\author{Manher Jariwala}\affiliation{Department of Physics, Boston University,  Boston, MA, 02215, USA} 
\author{Xochith Herrera}\affiliation{Department of Science Education, California State University Chico, Chico, CA, 95929, USA} 
\author{Eleanor W. Close}\affiliation{Department of Physics, Texas State University,  San Marcos, TX 78666, USA} 
\author{Ben Van Dusen}\affiliation{Department of Science Education, California State University Chico, Chico, CA, 95929, USA} 
\begin{abstract}
Research-based assessments (RBAs), such as the Force Concept Inventory, have played central roles in many course transformations from traditional lecture-based instruction to research-based teaching methods.  In order to support instructors in assessing their courses, the online Learning About STEM Student Outcomes (LASSO) platform simplifies administering, scoring, and interpreting RBAs. Reducing the barriers to using RBAs will support more instructors in objectively assessing the efficacy of their courses and, subsequently, transforming their courses to improve student outcomes. The purpose of this study was to investigate the extent to which RBAs administered online and outside of class with the LASSO platform provided equivalent data to traditional paper and pencil tests administered in class. Research indicates that these two modes of administering assessments provide equivalent data for graded exams that are administered in class. However, little research has focused on ungraded (low-stakes) exams that are administered outside of class. We used an experimental design to investigate the differences between these two test modes. Results indicated that the LASSO platform provided equivalent data to paper and pencil tests.
 \end{abstract}

\maketitle
\section{Introduction}
Research-based assessments (RBAs) are often used to both develop and disseminate research-based teaching methods that improve student outcomes. Subsequently, RBAs are the focus of many influential publications in physics education research, such as Hake's \citep{Hake1998} comparison of traditional and interactive-engagement courses. And, a large increase in the number of RBAs in physics education research coincided with a dramatic increase in the collaboration in the PER community \citep{Anderson2017}. Because of these successes, educators are interested in using RBAs. However,  \citet{Madsen2016} found that many instructors want support in choosing appropriate assessments, administering and scoring the assessments, and interpreting the results of their assessments. To address these needs the Learning Assistant Alliance developed the LASSO platform to host and administer RBAs online \cite{LASSO}. Hosting the RBAs online meets instructors' needs by allowing for the RBAs to be administered outside of class, to be promptly and automatically scored, and for instructors to be provided with a summary report to help interpret the results. 
\par 	Extensive research has investigated the differences between computer based tests (CBTs) and pencil and paper tests (PPTs). Meta analysis of the literature has revealed that there is no systematic difference in scores between these two modes of administering tests \citep{Wang2007,Wang2007a}. However, the studies in these meta analyses were conducted at the K-12 level, and most had the CBT being administered in class.  Because the LASSO platform is designed to administer RBAs outside of class in order to free up class time, the results of this earlier work may not apply to the LASSO platform.  
\par In a similar study to this one, \citet{Bonham2008} conducted research in college astronomy courses and administered assessments both online outside of class and in class. Bonham and colleagues had students complete both a locally-made concept inventory and a research-based attitude survey. The students were randomly assigned to two conditions with either the concept inventory done in class and the attitude survey done outside of class via an online system or the reverse. A matched sample was then drawn from the students who completed the surveys. They concluded that there was no significant difference between CBT and PPT data collection. However, a close analysis of their results revealed that there was a small but meaningful difference in the data and that the study did not have a sufficient sample size to rule out any meaningful differences; their study was underpowered. Their results indicated that the online concept inventory scores were 6\% higher than the in class scores, which was an effect size of  approximately 0.30. While this is a small difference, lecture-based courses often have raw gains below 20\% and a 6\% difference would skew comparisons between data collected with CBT and PPT modes. Therefore, it is not clear from the prior literature that low-stakes tests provide similar data when collected in class with PPTs or outside of class with CBTs.

\section{Research Question}
\par	The purpose of the present study was to inform if data collected outside of class with CBTs is consistently different than data collected in class with PPTs. In pursuit of this purpose we asked:

\begin{itemize}
\item \emph{To what extent does the online administration of RBAs outside of class using the LASSO platform provide equivalent data to the in-class administration of RBAs using PPTs?}  
\end{itemize}

If the LASSO platform provided equivalent data to paper-based administration, then it represents a much simpler entry point for instructors to begin assessing and transforming their own courses because it addresses many of the instructors' needs that \citet{Madsen2016} identified. A second major benefit of the widespread use of the LASSO system is that it automatically aggregates all of the data and makes this data available for research. The size and variety of this data allows for investigations that would have been underpowered if conducted at only a few institutions or lacking generalizability if only conducted in a few courses at a single institution. 

\section{Methods and Design}
The study was conducted at a medium-sized regional university in the United States. The data was collected in three different introductory physics courses: algebra-based mechanics, calculus-based mechanics, and calculus-based electricity and magnetism course (E\&M). The data was collected across two semesters from a total of 25 different sections in the three courses. 

\par	The study used a between-groups experimental design. Stratified random sampling created two random samples within each section with similar gender, race, and honors status makeups. One sample was assigned to complete a concept inventory online outside of class using the LASSO system and an attitudes survey in class using paper and pencil, which is Group 1 in Figure 1. We referred to this group as the CBT condition because our focus in this study was on the concept inventories. The other sample was assigned the concept inventory in class and the attitude survey online outside of class, which we referred to as the PPT condition. Within each course both groups completed the in-class assessment the same class period and had the same deadline to complete the online assessments. Assessments were collected at the beginning and end of the semester. Paper-based assessments were collected by the instructors, scanned using automated equipment, and uploaded to the LASSO system. Students' assessment data was downloaded from the LASSO system and combined with student grades and demographics data provided by the university. Only students who were assigned to a condition at the beginning of the semester were included in the data analysis. This resulted in a the total sample of 1,310 students. No other filters were applied to the data.
\par 	Students in both mechanics courses were assigned the Force Concept Inventory. Students in the E\&M course were assigned the Conceptual Survey of Electricity and Magnetism. Both assessments were scored on a 0-100\% scale. Students in all of the courses were assigned the Colorado Learning Attitudes about Science Survey. 

\begin{figure}
\includegraphics[width=1\columnwidth]{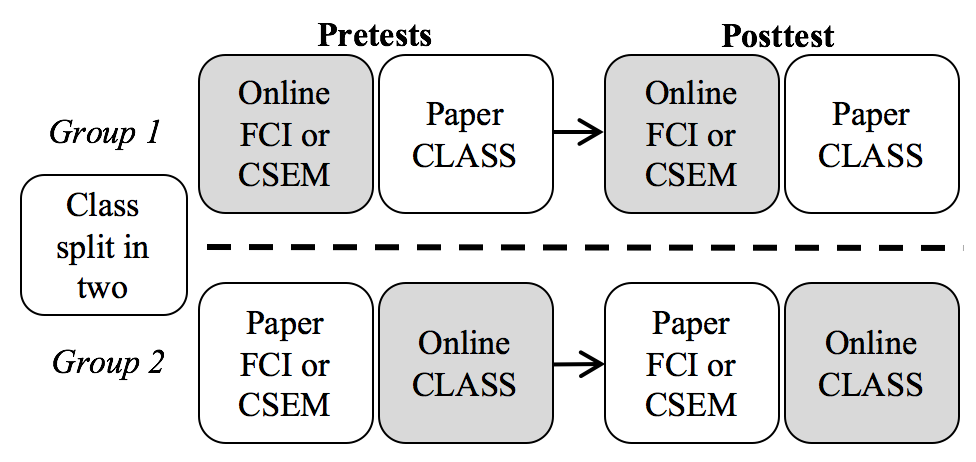}
\caption{Design of the research conditions.}
\end{figure}

\par	Completion rates for the PPT condition were 94\% for the pretest and 74\% for the posttest and for the CBT were 68\% for the pretest and 54\% for the posttest. Participation rates for this data  are discussed in detail by \citet{Jariwala2017}. Missing data was replaced using Hierarchical Multiple Imputation (HMI) with the mice package in R. HMI is a form of multiple imputation (MI) that takes into account the fact that students were nested in different courses and and that their performance may have been related to the course they were in. MI addresses missing data by imputing the missing data \emph{m} times to create \emph{m} complete data sets, analyzing each data set independently, and combining the \emph{m} results using standard methods \cite{Dong2013}. Our MI produced \emph{m}=10 complete data sets. Multiple imputation is preferable to list-wise deletion because it maximizes the statistical power of the study \cite{Dong2013} and has the same basic assumptions. 
\par 	We used the HLM 7 software package to analyze the data using Hierarchical Linear Modeling (HLM). HLM was necessary because of the nested structure of the data: students (level 1) were nested within course sections (level 2). HLM works by creating a linear model for each course and then combining those models to produce an overall hierarchical linear model. One indicator of the need for conducting HLM, as opposed to ordinary least squares (OLS) analysis, is the intraclass correlation coefficient (ICC), which informs how much of the variation in the data is at the student level versus at the course level by dividing the course level variance by the total variance. Thus, the ICC indicates how much information would be ignored by OLS analysis. The ICC for the pretest was 10.2\% and the ICC for the posttest was 19.8\%. No rule of thumb exists for a cutoff that requires HLM, but ICCs as low as 5\% can have large effects on the resultant analysis. 

%Table1: 
\begin{table*}[t]
\caption{Hierarchical Linear Models}
\begin{tabular}{p{2cm}ccp{.2cm}ccp{.2cm}ccp{.3cm}ccp{.2cm}ccp{.2cm}cc}
\hline \hline
\rule{0pt}{3ex}&\multicolumn{17}{c}{Fixed Effects with Robust SE}\\ \cline{2-18}
\rule{0pt}{3ex}& \multicolumn{8}{c}{Pretest}&&\multicolumn{8}{c}{Posttest}\\ \cline{2-9} \cline{11-18}
\rule{0pt}{3ex} &\multicolumn{2}{c}{\underline{Model E1}}&&\multicolumn{2}{c}{\underline{Model E2}}&&\multicolumn{2}{c}{\underline{Model E3}}&&\multicolumn{2}{c}{\underline{Model O1}}&&\multicolumn{2}{c}{\underline{Model O2}}&&\multicolumn{2}{c}{\underline{Model O3}}\\
			&b&\emph{p}&&b&\emph{p}&&b&\emph{p}&&b&\emph{p}&&b&\emph{p}&&b&\emph{p}\\
For Intercept	&&&&&&&&&&&&&&&&&\\
~~Intercept	&30.95&\textless0.001	&&31.15&\textless0.001	&&26.9&\textless0.001	&&44.0&\textless0.001	&&44.0&\textless0.001	&&34.3&\textless0.001\\
~~E\&M		&-&-&			&-&-&			&2.84&0.114&			&-&-&			&-&-&			&6.98&0.006\\
~~Calculus	&-&-&			&-&-&			&9.58&\textless0.001&	&-&-&			&-&-&			&20.7&\textless0.001\\
For CBT	&&&&&&&&&&&&&&&&&\\
~~Intercept	&-&-&			&-0.40&0.63&		&0.09&0.93&			&-&-&			&-0.02&0.99&		&-0.21&0.87\\
~~E\&M		&-&-&			&-&-&			&-1.32&0.57&			&-&-&			&-&-&			&2.31&0.46\\
~~Calculus	&-&-&			&-&-&			&-0.48&0.79&			&-&-&			&-&-&			&-0.98&0.69\\
&&&&&&&&&&&&&&&&&\\
&\multicolumn{17}{c}{Random Effects}\\ \cline{2-18}
\rule{0pt}{3ex}& \multicolumn{8}{c}{Pretest}&&\multicolumn{8}{c}{Posttest}\\ \cline{2-9} \cline{11-18}
\rule{0pt}{3ex}&\multicolumn{2}{c}{\underline{Model E1}}&&\multicolumn{2}{c}{\underline{Model E2}}&&\multicolumn{2}{c}{\underline{Model E3}}&&\multicolumn{2}{c}{\underline{Model O1}}&&\multicolumn{2}{c}{\underline{Model O2}}&&\multicolumn{2}{c}{\underline{Model O3}}\\
Intercept Var&\multicolumn{2}{c}{20.8}&&\multicolumn{2}{c}{22.3}&&\multicolumn{2}{c}{3.4}&&\multicolumn{2}{c}{83.7}&&\multicolumn{2}{c}{84.8}&&\multicolumn{2}{c}{6.68}\\
 Condition Var&\multicolumn{2}{c}{-}&&\multicolumn{2}{c}{0.41}&&\multicolumn{2}{c}{0.48}&&\multicolumn{2}{c}{-}&&\multicolumn{2}{c}{0.95}&&\multicolumn{2}{c}{2.19}\\
\hline \hline

\end{tabular}
\end{table*}

\par 	Following from our purpose to test if CBT and PPT conditions provided similar information, we tested the impact of the condition on the pretest and on the posttest in two different sets of models. Initially, we built these models to control for course grade, gender, and under-represented minority status because these variables are known to relate to performance on RBAs. However, these controls provided no additional information relevant to the role of the CBT condition in the models. For brevity, we excluded these variables from the models presented in this article. The final pretest model consisted of pretest score as the outcome variable, CBT condition as the predictor variable and level 2 variables for the calculus-based mechanics course and the E\&M course. Variables were only necessary for two of the three courses as the intercept represented the third course: algebra-based mechanics. The final posttest model consisted of posttest score as the outcome variable, CBT condition as the predictor variable, and level 2 variables for the courses. The models were built in three steps: (1) no predictors, (2) then add level 1 predictors, and (3) then add level 2 predictors. This three-step process informed how much additional information was being explained by the addition of the new predictors in each step as indicated by a reduction in the variance for that variable. One distinction between HLM and OLS regression is that in OLS the addition of variables always reduces the unexplained variance, whereas in HLM ``if a truly nonsignificant variable enters the model, it is mathematically possible under maximum likelihood to observe a slight increase in the residual variance'' \citep[p.~150]{raudenbush2002}. Once the models were constructed we used the hypothesis testing function in the HLM 7 software to generate predicted values for each of the courses pre- and posttest scores for both conditions with 95\% confidence intervals to inform the size and reliability of any differences between conditions.

\begin{figure}
\includegraphics[width=1\columnwidth]{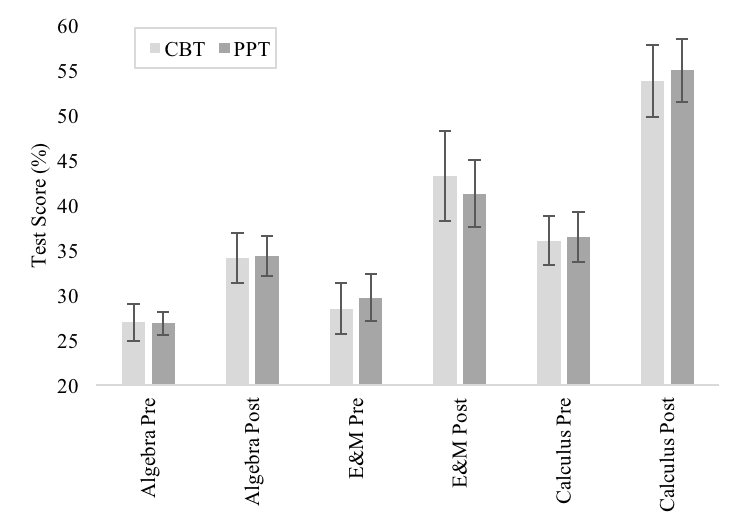}
\caption{Predicted Mean Scores with 95\% CIs.}
\end{figure}

\section{Results}
The first pretest model, Model E1 in Table 1, indicated that all students on average had a pretest score of 30.95, that this mean was statistically reliably different from zero (p\textless0.001) and that there was a level 2 variance of 20.8 for the intercept. The addition of CBT condition to the model, Model E2, had no impact on the model. The intercept for CBT condition was very small, -0.40, and was not statistically reliably different from 0 (p=0.63). This result indicated that whether or not students completed the pretest online or in class had no meaningful impact on their score. Furthermore, the Intercept variance increased from 20.8 to 22.3 when CBT condition was added to the model, indicating that CBT condition was a truly nonsignificant variable. The addition of the level 2 variables, Model E3, provided further evidence that CBT condition was nonsignificant. For algebra-based mechanics students there was effectively no difference, only 0.09. For calc-based mechanics students there was an extremely small negative difference, -0.39 (calculated by adding the intercept, 0.09, and the Calculus value, -0.48). For the E\&M students there was a small difference -1.23. None of these values were statistically reliable (p\textgreater \textgreater 0.05). Furthermore, while the addition of the level 2 variables greatly reduced the unexplained variance for the Intercept they increased the unexplained variance for CBT condition, indicating that CBT condition was truly nonsignificant on the pretest.
\par 	The posttest models revealed similar trends to the pretest models with regards to CBT condition. The addition of the CBT condition variable in Model O2 had no meaningful relationship to students' scores, -0.02, and increased the variance for the Intercept. Adding the level 2 variables to the model, Model O3, provided further evidence that CBT condition was not meaningfully related to students' posttest scores. For algebra-based mechanics students there was effectively no difference, -0.21. For calc-based mechanics students there was a small negative difference, -1.19. For the E\&M students there was a small positive difference, 2.10. None of these values were statistically reliable (p\textgreater \textgreater0.05). Furthermore, the addition of the level 2 variables increased the CBT condition predictor's variance from 0.95 to 2.19, indicating that CBT condition was truly nonsignificant on the posttest.
\par The largest predicted effect of CBT condition was on the posttest for E\&M students. This predicted effect bordered on being large enough to be meaningful because it indicated a 2.2 points higher posttest score for students doing the CBT and the overall predicted gain for the E\&M students was only 11.6 points. However, the pre- and posttest across the three courses created six total measurements of the predicted effect for CBT condition, shown in Figure 1; in three of those measurements the effect was nearly zero, in one it was positive, and in two it was negative. In addition to these inconsistencies in all six comparisons across CBT and PPT conditions there was large overlap in the 95\% confidence intervals, indicating that the differences were not  reliably different. 

\section{Conclusion and Implications}
The differences between students' predicted grades for the PPT and CBT conditions were very small, did not consistently favor one condition over the other, and were not statistically reliable by any metric. The nonsignificance of CBT condition was strongly indicated by the addition of CBT condition to the models  increasing the variance in every case. This evidence led us to conclude that there was no meaningful difference in test scores on low-stakes RBAs between students who completed the exam in class using paper and pencil and those who completed the exam outside of class using the online LASSO platform. This similarity indicates that instructors and researchers can use the LASSO platform to collect valuable and normalizable information about the impacts of their courses without concerns about the legitimacy of comparing that data to prior research that was collected with paper and pencil tests. Collecting data with the LASSO system can greatly reduce the barriers to instructor's use of RBAs since instructors do not need to dedicate class time to collect the data or their own time to sort, scan, and analyze the data. It is important to note, however, that instructors do need to make some effort to motivate their students to complete the online assessments. We have found that by making announcements in class, sending out email reminders, and giving credit to students who complete the RBAs instructors can achieve similar participation rates on CBT assessments as on PPT assessments \cite{Jariwala2017}. Participation rates in courses that do not use these practices can be quite low. Our hope is that reducing the barriers to RBA use will lead more instructors to assess the efficacy of their courses and, subsequently, to adopt research-based teaching practices that support student success.

\section{Limitations}
The results of this study are only applicable to asking students to complete one low-stakes assessment online at the beginning and end of the course. Excessive measurement would likely decrease student participation, performance, and data quality. Higher stakes would likely incentivize using additional materials not available for tests administered in class. Comparisons of CBT and PPT administered assessments may also be impacted by missing data. Our use of HMI should have mitigated any impacts of missing data, but studies that use list-wise deletion to address missing data may have different results because of the non-ignorable nature of missing RBA data. It is also possible that the institution where the study was conducted and the populations involved in the study are not representative of physics students or courses broadly. However, the study included three different courses that included both calculus-based and algebra-based physics sequences, which supports the results applying to many populations of students. 
\par This work was funded in part by NSF-IUSE Grant No. DUE-1525338 and is Contribution No. LAA-048 of the Learning Assistant Alliance.
\bibliography{gender}

\end{document}